\documentclass[preprint, superscriptaddress, showpacs,preprintnumbers,amsmath,amssymb,prb]{revtex4}
\usepackage{graphicx}

\begin{document}

\thispagestyle{empty}

\title{Van der Waals and Casimir interactions between
two graphene sheets}

\author{
G.~L.~Klimchitskaya
}
\affiliation{Central Astronomical Observatory
at Pulkovo of the Russian Academy of Sciences,
St.Petersburg, 196140, Russia}

\author{
 V.~M.~Mostepanenko
}
\affiliation{Central Astronomical Observatory
at Pulkovo of the Russian Academy of Sciences,
St.Petersburg, 196140, Russia}

\begin{abstract}
The thermal free energy and pressure of dispersion interaction
between two graphene sheets described by the Dirac model are
calculated using the Lifshitz formula with reflection coefficients
expressed via the polarization tensor. The obtained results for
a pristine graphene are found to be in agreement with computations
using Coulomb coupling between density fluctuations. For a
graphene with nonzero mass gap parameter a qualitatively different
behavior for the free energy and pressure is obtained.
The Lifshitz formula with reflection coefficients expressed
via the polarization tensor is used as a test for different
computational approaches proposed in the literature for modeling
the response function and conductivity of graphene at both zero
and nonzero temperature.
\end{abstract}

\maketitle

\section{Introduction}

The van der Waals and Casimir interactions, which are known under
the generic name of dispersion forces,\cite{1} are caused by the
vacuum and thermal fluctuations of the electromagnetic field.
At shortest separations of a few nanometers dispersion forces are
usually referred to as the {\it van der Waals} forces.
At larger separations, when the relativistic retardation becomes
important, it is customary to speak about the {\it Casimir} forces.
In the last few years the fluctuation induced forces attracted
much attention in the literature\cite{2,3,4} due to their
prospective applications in both fundamental physics and
nanotechnology. Specifically, a lot of experiments has been
performed\cite{5,6,7} on measuring dispersion forces between
metallic, dielectric and semiconductor surfaces spaced at
separations from a few tens to a few hundreds nanometers.

Recently, special attention has been directed to carbon
nanostructures, such as one-atom-thick graphene sheets,
carbon nanotubes, fullerenes etc. which possess unique
mechanical, electrical and optical properties.\cite{8,9}
These properties appear to be particularly promising to
provide the basis for future carbon-based nanoelectronics.
Keeping in mind that elements of nano- and
microelectromechanical devices are separated by distances
of the order of tens or hundreds nanometers, the dispersion
forces acting between them are gaining in importance.
For this reason, a lot of papers has been devoted to
calculations of the van der Waals and Casimir forces
between two carbon nanostructures and between a carbon
nanostructure and a regular material
body.\cite{10,11,12,13,14,15,16,17,18,19,20,21,22,23}
Particular
attention has been given also the the Casimir-Polder
interaction of different atoms and molecules with carbon
nanostructures.\cite{24,25,26}

The many and varied formalisms were applied to calculate the
van der Waals and Casimir forces between two graphene sheets.
Here we center our attention on the approaches consistent with
the most realistic {\it Dirac} model which assumes the linear
dispersion relation of the graphene bands at low energy\cite{9}
(there is also the so-called {\it hydrodynamic} model of
graphene\cite{13,14,27} which does not take this property into
account). Specifically, we compare the computational results
obtained using the density fluctuation approach and the random
phase approximation,\cite{10,16,20,21} and using the conductivity
of graphene modeled as a combination of Lorentz-type oscillators
without account\cite{18} and with account\cite{22} of spatial
dispersion. It should be emphasized that the most straightforward
formalism for calculation of the van der Waals and Casimir forces
between graphene and different substances within the Dirac model
is based on the Lifshitz theory and exploits the reflection
coefficients of the electromagnetic oscillations on graphene
expressed in terms of the polarization tensor in
(2+1)-dimensional space-time.\cite{15,17}
Using this formalism, the van der Waals and Casimir interactions
between a graphene sheet and an ideal metal plane,\cite{15,17}
and between a graphene sheet and different atoms\cite{26} and
plates made of various real materials\cite{23} were computed.
The case of graphene-graphene interaction remained, however,
unexplored within this calculation approach.

The present paper is devoted to calculation of the
graphene-graphene thermal van der Waals and Casimir interactions
using the Lifshitz theory and the Dirac model for electronic
properties of graphene. The reflection coefficients of the
electromagnetic oscillations on graphene are expressed directly
through the polarization tensor without recourse to the concept
of dielectric permittivity. All calculations are performed for
both gapless pristine graphene and graphene sheets characterized
by some nonzero mass gap parameter. Note that the Dirac-type
excitations in graphene become massive under the influence of
electron-electron interaction, substrates, defects of structure
and some other effects.\cite{9,28,29,30,31}
We calculate both the van der Waals and Casimir free energy
per unit area and
pressure as functions of separation between the
graphene sheets and the van der Waals and Casimir pressure as a function of
temperature. For a gapless graphene our results are in agreement
with computations using Coulomb coupling between density
fluctuations with subsequent thermal average.\cite{16}
The latter approach is in fact equivalent\cite{16} to the
nonretarded limit of the Lifshitz formula with the polarization
of an isolated graphene sheet described in the random phase
approximation. It was also shown,\cite{16} that retardation
effects are of only minor importance for graphene.
For graphene sheets with nonzero mass gap parameter we obtain
a qualitatively different behavior for the free energy and
pressure,
as compared to the pristine graphene. In this case the character
of force depends on the relationship between the mass gap
parameter and the temperature. We provide a discussion concerning
the comparison of our results with other results obtained in the
literature for graphene-graphene van der Waals and Casimir
interactions using the Dirac model.\cite{10,18,20,21,22}

The paper is organized as follows. In Sec.~II we begin with the
Lifshitz formula containing the reflection coefficients derived
using the Dirac model. Then we present our results for the free
energy of graphene-graphene van der Waals and Casimir
interactions per unit area and pressure as functions of separation
 and temperature. Section~III contains the comparison of
our results with the results by others obtained at both zero and
nonzero temperature. In Sec.~IV the reader will find our
conclusions and discussion. For simplicity in comparisons with
the classical limit, we preserve the fundamental constants
(the Planck constant $\hbar$, the velocity of light $c$ and
the Boltzmann constant $k_B$) in all mathematical expressions.

\section{Dispersion interaction of two graphene sheets described
by the Dirac model}

The free energy of the van der Waals and Casimir
interactions per unit area of two parallel graphene sheets
separated by a distance $a$ at thermal equilibrium with an
environment at temperature $T$ is given by the Lifshitz
formula\cite{1,2,3,4,5,6,7}
\begin{equation}
{\cal F}(a,T)=\frac{k_BT}{8\pi a^2}\sum_{l=0}^{\infty}
{\vphantom{\sum}}^{\prime}\int_{\zeta_l}^{\infty}\!\!\!\!y\,dy
\left\{\ln\left[1-r_{\rm TM}^{2}(i\zeta_l,y)
e^{-y}\right]
 +\ln\left[1-r_{\rm TE}^{2}(i\zeta_l,y)
e^{-y}\right]\right\}.
\label{eq1}
\end{equation}
\noindent
Here,
$\zeta_l$ are the dimensionless Matsubara frequencies connected
with the dimensional ones $\xi_l=2\pi k_BTl/\hbar$
(where $l=0,\,1,\,2,\,\ldots$) by the
equality $\zeta_l=\xi_l/\omega_c$ with $\omega_c=c/(2a)$.
The dimensionless variable $y$ is connected with the
magnitude of the projection of the wave vector on the plane of
graphene, $k_{\bot}$, by the equality
$y=2a(k_{\bot}^2+\xi_l^2/c^2)^{1/2}$.
 The prime near the summation sign means
that the term with $l=0$ should be taken with a factor 1/2.

In the framework of the Dirac model at $T\neq 0$
the reflection coefficients on graphene for two independent
polarizations of the electromagnetic field, transverse magnetic
(TM) and transverse electric (TE), were found in Ref.~\cite{17}.
In terms of our dimensionless variables they are given
by\cite{23,26}
\begin{eqnarray}
&&
r_{\rm TM}(i\zeta_l,y)=
\frac{y\tilde{\Pi}_{00}}{y\tilde{\Pi}_{00}+
2(y^2-\zeta_l^2)},
\label{eq2} \\
&&
r_{\rm TE}(i\zeta_l,y)=
-\frac{(y^2-\zeta_l^2)\tilde{\Pi}_{tr}-
y^2\tilde{\Pi}_{00}}{(y^2-\zeta_l^2)(\tilde{\Pi}_{tr}
+2y)-y^2\tilde{\Pi}_{00}},
\nonumber
\end{eqnarray}
\noindent
where the dimensionless components of the polarization tensor
in (2+1)-dimensional space-time are expressed as
$\tilde{\Pi}_{00,tr}={2a}{\Pi}_{00,tr}/{\hbar}$
through the dimensional ones
and trace stands for the sum of spatial components
$\Pi_{1}^{\,1}$ and $\Pi_{2}^{\,2}$.

The explicit expression for the 00-component of the polarization
tensor for graphene with a nonzero mass gap parameter $\Delta$
but zero chemical potential can be written in the
form\cite{17,23,26}
\begin{eqnarray}
&&
\tilde{\Pi}_{00}(i\zeta_l,y)=8\alpha(y^2-\zeta_l^2)
\int_{0}^{1}dx\frac{x(1-x)}{\left[{\tilde{\Delta}}^2+
x(1-x)f(\zeta_l,y)\right]^{1/2}}
+\frac{8\alpha}{{\tilde{v}}_F^2}\int_{0}^{1}dx
\label{eq3} \\
&&
~\times
\left\{\vphantom{\frac{{\tilde{\Delta}}^2+\zeta_l^2x(1-x)}{\left[{\tilde{\Delta}}^2+
x(1-x)f(\zeta_l,y)\right]^{1/2}}}
\frac{\tau}{2\pi}\ln\left[1+2\cos(2\pi lx)e^{-g(\tau,\zeta_l,y)}
+e^{-2g(\tau,\zeta_l,y)}\right]
-\frac{\zeta_l}{2}(1-2x)
\frac{\sin(2\pi lx)}{\cosh{g(\tau,\zeta_l,y)}+
\cos(2\pi lx)}
\right.
\nonumber \\
&&~
\left.
+\frac{{\tilde{\Delta}}^2+\zeta_l^2x(1-x)}{\left[{\tilde{\Delta}}^2+
x(1-x)f(\zeta_l,y)\right]^{1/2}}\,
\frac{\cos(2\pi lx)+e^{-g(\tau,\zeta_l,y)}}{\cosh{g(\tau,\zeta_l,y)}+
\cos(2\pi lx)}
\right\}.
\nonumber
\end{eqnarray}
\noindent
Here, $\alpha=e^2/(\hbar c)$ in the fine-structure constant,
$\tilde{\Delta}=\Delta/(\hbar\omega_c)$ is the
dimensionless mass gap parameter,
$\tilde{v}_F=v_F/c\approx 1/300$ is the dimensionless Fermi velocity,
 and $\tau=2\pi T/T_{\rm eff}=4\pi ak_BT/(\hbar c)$, where
 $T_{\rm eff}$ is the so-called {\it effective temperature}.
The dimensionless functions $f$ anf $g$ contained in
Eq.~(\ref{eq3}) are
defined as
\begin{eqnarray}
&&
f(\zeta_l,y)={\tilde{v}}_F^2y^2+(1-{\tilde{v}}_F^2)\zeta_l^2,
\label{eq4} \\
&&
g(\tau,\zeta_l,y)=\frac{2\pi}{\tau}
\left[{\tilde{\Delta}}^2+x(1-x)f(\zeta_l,y)\right]^{1/2}.
\nonumber
\end{eqnarray}
\noindent
The explicit expression for the trace of the polarization
tensor  is given
by\cite{17,23,26}
\begin{eqnarray}
&&
\tilde{\Pi}_{tr}(i\zeta_l,y)=8\alpha[y^2+f(\zeta_l,y)]
\int_{0}^{1}dx\frac{x(1-x)}{\left[{\tilde{\Delta}}^2+
x(1-x)f(\zeta_l,y)\right]^{1/2}}
+\frac{8\alpha}{{\tilde{v}}_F^2}\int_{0}^{1}dx
\label{eq5} \\
&&
~\times
\left\{\vphantom{\frac{{\tilde{\Delta}}^2+\zeta_l^2x(1-x)}{\left[{\tilde{\Delta}}^2+
x(1-x)f(\zeta_l,y)\right]^{1/2}}}
\frac{\tau}{2\pi}\ln\left[1+2\cos(2\pi lx)e^{-g(\tau,\zeta_l,y)}
+e^{-2g(\tau,\zeta_l,y)}\right]
\right.
\nonumber \\
&&~
-\frac{\zeta_l(1-2{\tilde{v}}_F^2)}{2}(1-2x)
\frac{\sin(2\pi lx)}{\cosh{g(\tau,\zeta_l,y)}+
\cos(2\pi lx)}
\nonumber \\
&&~
\left.
+\frac{{\tilde{\Delta}}^2+x(1-x)[(1-{\tilde{v}}_F^2)^2\zeta_l^2-
{\tilde{v}}_F^4y^2]}{\left[{\tilde{\Delta}}^2+
x(1-x)f(\zeta_l,y)\right]^{1/2}}\,
\frac{\cos(2\pi lx)+e^{-g(\tau,\zeta_l,y)}}{\cosh{g(\tau,\zeta_l,y)}+
\cos(2\pi lx)}
\right\}.
\nonumber
\end{eqnarray}

The Lifshitz formula for the pressure of dispersion interaction
between two parallel graphene sheets takes the form
\begin{equation}
P(a,T)=-\frac{k_BT}{8\pi a^3}\sum_{l=0}^{\infty}
{\vphantom{\sum}}^{\prime}\int_{\zeta_l}^{\infty\!\!\!\!}y^2\,dy
\left\{\left[r_{\rm TM}^{-2}(i\zeta_l,y)
e^{y}-1\right]^{-1}+\left[r_{\rm TE}^{-2}(i\zeta_l,y)
e^{y}-1\right]^{-1}\right\}.
\label{eq6}
\end{equation}

\begin{figure}[b]
\vspace*{-3cm}
\centerline{\hspace*{1cm}
\includegraphics{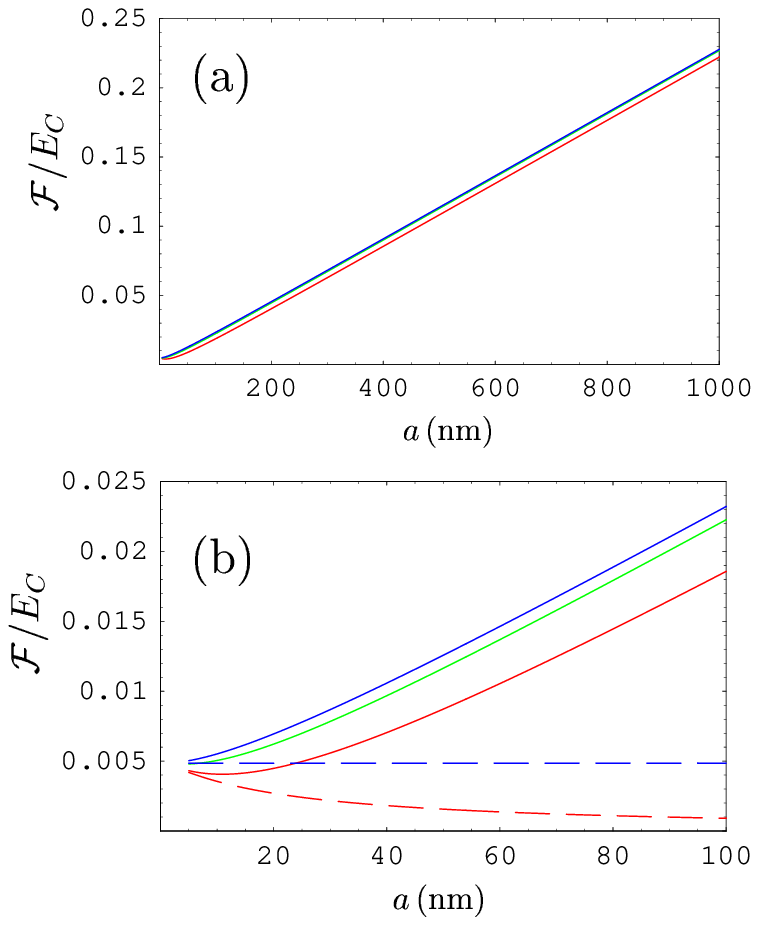}
}
\vspace*{-15cm}
\caption{\label{f1}(Color online)
The normalized van der Waals and Casimir free energy
of graphene-graphene interaction per unit area
as a function of separation
(a) from 5 to 1000\,nm and (b) from 5 to 100\,nm.
The solid lines from bottom to top are for
$T=300\,$K and the
mass gap parameter $\Delta=0.1\,$eV, 0.05\,eV, and 0\,eV,
respectively.
The bottom and top long-dashed lines are plotted
at $T=0$ for $\Delta=0.1\,$eV and 0\,eV,
respectively.
}
\end{figure}

We begin with computations of the free energy of dispersion
interaction between two graphene sheets using
Eqs.~(\ref{eq1})--(\ref{eq5}). In Fig.~\ref{f1}(a,b) we plot
the computational results for the free energy per unit area
${\cal F}$ at $T=300\,$K normalized to the Casimir energy
between two ideal metal planes at zero temperature,
$E_C(a)=-\pi^2\hbar c/(720a^3)$, in the separation region
(a) from 5 to 1000\,nm and (b) on an enlarged scale from 5
to 100\,nm. The bottom and top solid lines in Fig.~\ref{f1}(a)
show the ratio ${\cal F}/E_C$ at $T=300\,$K for graphene
sheets with the mass gap parameter $\Delta=0.1\,$eV and 0\,eV,
respectively. In Fig.~\ref{f1}(b) the quantity
${\cal F}/E_C$ at $T=300\,$K is shown by the solid lines
from bottom to top for $\Delta=0.1\,$eV, 0.05\,eV and 0\,eV,
respectively, and at $T=0\,$K by the bottom and top long-dashed
lines for $\Delta=0.1\,$eV and 0\,eV, respectively.
As can be seen in Fig.~\ref{f1}(a,b), at short separations
the free energy of graphene-graphene dispersion interaction is
much smaller than the Casimir interaction between two ideal
metal planes, but becomes relatively large with increasing
separation distance. The nonzero gap parameter depending on its
value exerts some influence on the free energy.
{}From Fig.~\ref{f1}(b) it is seen that already at short
separations from 10 to 20\,nm the  computational results at
$T=300\,$K differ considerably from those at $T=0\,$K.
This means that for graphene thermal effects should be taken
into account not only for the Casimir force, but for the
nonrelativistic van der Waals force as well.\cite{16}

\begin{figure}[b]
\vspace*{-3cm}
\centerline{\hspace*{1cm}
\includegraphics{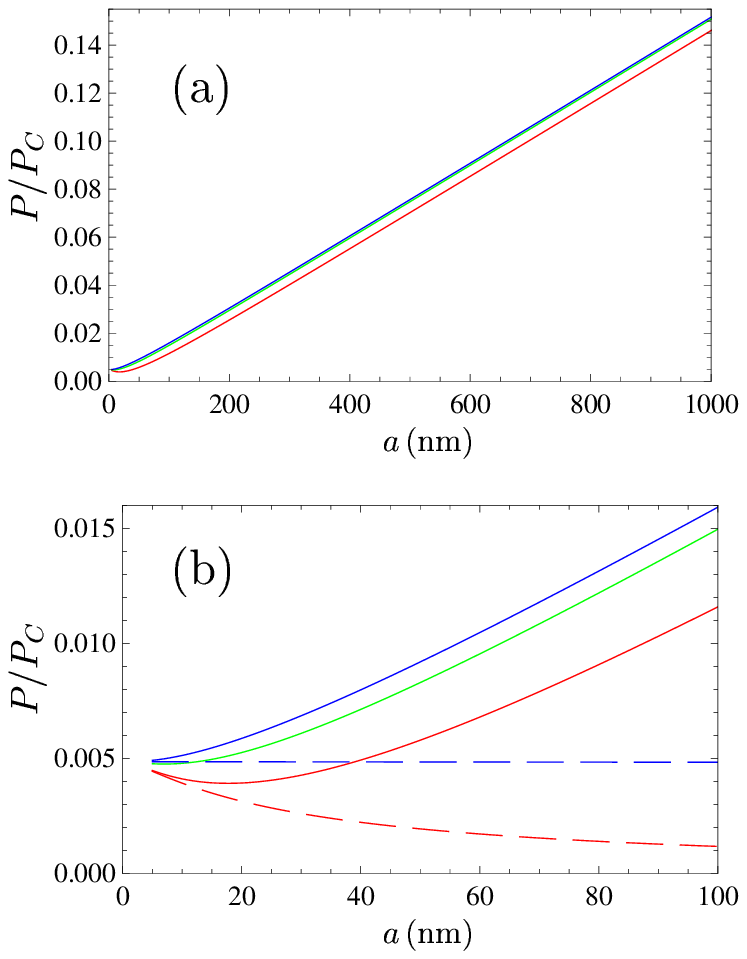}
}
\vspace*{-15cm}
\caption{\label{f2}(Color online)
The normalized van der Waals and Casimir pressure
of graphene-graphene interaction
as a function of separation
(a) from 5 to 1000\,nm and (b) from 5 to 100\,nm.
The solid lines from bottom to top are for
$T=300\,$K and the
mass gap parameter $\Delta=0.1\,$eV, 0.05\,eV, and 0\,eV,
respectively.
The bottom and top long-dashed lines are plotted
at $T=0$ for $\Delta=0.1\,$eV and 0\,eV,
respectively.
}
\end{figure}

The computations of the van der Waals and Casimir pressures
between two graphene sheets were performed using
Eqs.~(\ref{eq2})--(\ref{eq6}). In Fig.~\ref{f2}(a,b)
the computational results for the pressure $P$
 at $T=300\,$K normalized to the Casimir pressure
between two ideal metal planes at zero temperature,
$P_C(a)=-\pi^2\hbar c/(240a^4)$, are plotted
in the separation region
(a) from 5 to 1000\,nm and (b) on an enlarged scale from 5
to 100\,nm. The meaning of the solid and the long-dashed
lines is the same as in  Fig.~\ref{f1}(a,b).
{}From  Fig.~\ref{f2}(a,b) it can be concluded that the
pressure of graphene-graphene dispersion interaction possesses
all the same properties as discussed above in the case of free
energy. Specifically, at $T=300\,$K the thermal effect becomes
large enough at short separations of about 10--20\,nm and  its
role quickly increases with increasing separation depending on
the value of $\Delta$.

The top solid line in Fig.~\ref{f2}(b) related to the case of
pristine graphene ($\Delta=0$) within the separation region from
5 to 100\,nm is in agreement with the solid line in Fig.~2 of
Ref.~\cite{16} obtained using Coulomb couplings between density
fluctuations with subsequent thermal average.
As was shown,\cite{16} thermal effects
in the pressure of dispersion interaction
for graphene are
noticeable already in the van der Waals regime, i.e.,
at distances of tens of nanometers at room temperature $T=300\,$K.
{}From the comparison of the top solid and top long-dashed lines
in Fig.~\ref{f2}(b) one can calculate the relative thermal
correction to the Casimir pressure
\begin{equation}
\delta_TP(a)=\frac{P(a,T)-P_C(a)}{P_C(a)}
\label{eq7}
\end{equation}
\noindent
at different separations for a gapless graphene. Thus, at $a=10$,
20, 50, and 100\,nm one obtains $\delta_TP(a)=5.6$\%, 20.8\%,
89.4\%, and 228\%, respectively [remind that for customary
metallic and dielectric materials thermal correction becomes
large only at separations of a few micrometers comparable with
the so-called {\it thermal length}\cite{4}
$\hbar c/(2k_BT)$].

Our results show that for graphene sheets with a nonzero mass gap
parameter the thermal effect (with exception of only shortest
separations) becomes even larger. Thus, from
the comparison of the bottom solid line with the bottom long-dashed line
in Fig.~\ref{f2}(b) (graphene sheets with $\Delta=0.1\,$eV)
 at separations $a=10$,
20, 50, and 100\,nm we obtain $\delta_TP(a)=4.7$\%, 25.3\%,
200\%, and 893\%.
This makes feasible an observation of the thermal effect in the
dispersion interaction of two graphene sheets and even an estimation
of the mass gap parameter by the results of force measurements.

\begin{figure}[b]
\vspace*{-13cm}
\centerline{\hspace*{1cm}
\includegraphics{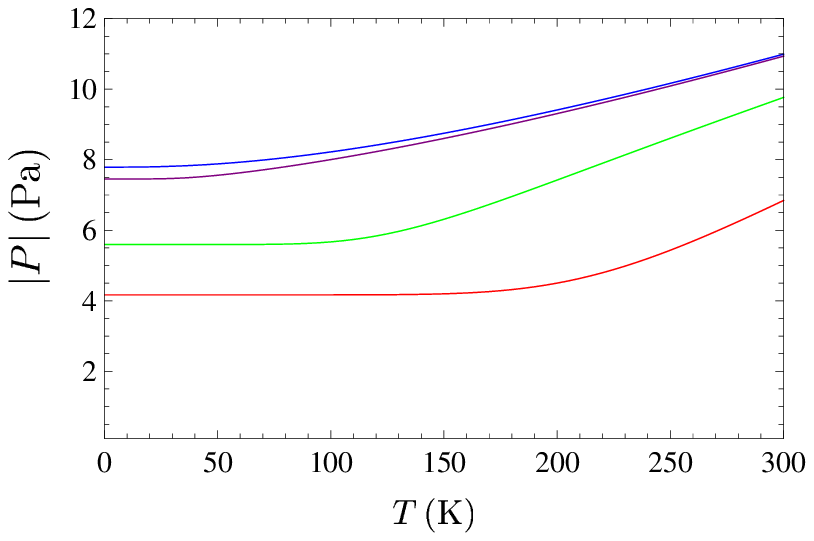}
}
\vspace*{-10cm}
\caption{\label{f3}(Color online)
The van der Waals and Casimir pressure
for graphene-graphene interaction
at $a=30\,$nm
as a function of temperature.
The lines from bottom to top are for
 the
mass gap parameter $\Delta=0.1\,$eV, 0.05\,eV,
0.01\,eV, and 0\,eV,
respectively.
}
\end{figure}

The role of the mass gap parameter in the temperature dependence
of the van der Waals and Casimir pressure between two graphene
sheets is illustrated in Fig.~\ref{f3}. Here, the separation
distance is fixed at $a=30\,$nm and the pressure magnitudes are
plotted as functions of temperature for
$\Delta=0.1\,$eV, 0.05\,eV, 0.01\,eV, and 0\,eV from bottom to
top, respectively. As can be seen in Fig.~\ref{f3}, for any
nonzero mass gap parameter there is some temperature region where
the pressure magnitude remains nearly constant when the
temperature increases. The larger is the mass gap parameter,
the wider is this temperature region. Thus, for
$\Delta=0.1\,$eV, 0.05\,eV, and 0.01\,eV the pressure magnitude
remains nearly constant (less than 1\% increase) up to $T=155\,$K,
95\,K, and 45\,K, respectively. Note that the same characteristic
dependence on $\Delta$ for dispersion interactions of graphene
with atomic systems and material plates made of dielectrics and metals was
found earlier.\cite{23,26}
The Casimir free energy remains nearly constant if the condition
$k_BT\ll\Delta$ is satisfied with a large safety margins.
Under the condition $\Delta\lesssim k_BT$ the thermal correction
becomes relatively large.

The results of numerical computations presented above can be
supplemented
by the asymptotic behaviors of the free energy and pressure at
large and short separations (high and low temperatures).
In the case of large separations (high temperatures) the
asymptotic behavior of the Casimir free energy and pressure is
determined by the zero-frequency contribution to the Lifshitz
formulas (\ref{eq1}) and (\ref{eq6}). Using the asymptotic
expressions for the reflection coefficients obtained
earlier,\cite{17,23} one arrives at the following Casimir free
energy per unit area in the large separation (high
 temperature) limit
 \begin{equation}
 {\cal F}(a,T)=-\frac{k_BT\zeta(3)}{16\pi a^2}\left[
 1-\frac{{\tilde{v}}_F^2\hbar c}{4\alpha ak_BT
 \ln\left(2\cosh\frac{\Delta}{2k_BT}\right)}\right],
 \label{eq8}
 \end{equation}
 \noindent
 where $\zeta(z)$ is the Riemann zeta function.
 Note that the second term in the square brackets of
 Eq.~(\ref{eq8})
 is small comparing with unity due to the smallness of
 $\tilde{v}_F$.
 In a similar way for the Casimir pressure at large separations
(high temperatures) it holds
\begin{equation}
 {P}(a,T)=-\frac{k_BT\zeta(3)}{8\pi a^3}\left[
 1-\frac{3{\tilde{v}}_F^2\hbar c}{8\alpha ak_BT
 \ln\left(2\cosh\frac{\Delta}{2k_BT}\right)}\right].
 \label{eq9}
 \end{equation}

The asymptotic expressions (\ref{eq8}) and (\ref{eq9}) are in
a good agreement with the results of numerical computations shown
in Figs.~\ref{f1}(a) and \ref{f2}(a). As an example, for
$\Delta=0$, $T=300\,$K the pressure values calculated using
Eq.~(\ref{eq9}) agree with computations in the limits of 1\% at
separations $a>370\,$nm. In the limits of 5\% the analytic and
computational results agree at $a>150\,$nm.
For graphene with nonzero mass gap parameter the
asymptotic expressions (\ref{eq8}) and (\ref{eq9}) become
applicable starting from larger separation distances.
Thus, for $\Delta=0.1\,$eV, $T=300\,$K the agreement between
analytic and numerical computations in the limits of 5\% is
achieved at $a\geq 800\,$nm.
Note that the first term on the right-hand side of Eq.~(\ref{eq9})
was obtained in Ref.~\cite{16} for a gapless graphene with
$\Delta=0$. The second terms in Eqs.~(\ref{eq8}) and (\ref{eq9})
provide first corrections to the previously obtained result and
generalize it to the case of graphene with a nonzero mass gap
parameter. It should be stressed that the main (first) terms
 on the right-hand side of Eqs.~(\ref{eq8}) and (\ref{eq9})
 correspond to the so-called {\it classical limit}\cite{32}
 because they do not depend on the Planck constant.

The case of two graphene sheets interacting via the nonthermal
van der Waals interaction is restricted to the shortest
separations from 1 to 3\,nm.
Here, using Eqs.~(\ref{eq1}) and (\ref{eq6}) at $T=0$, $\Delta=0$,
one obtains
\begin{equation}
E(a)={\cal F}(a,0)=-\frac{C}{a^3},\quad
P(a,0)=-\frac{3C}{a^4},
\label{eq10}
\end{equation}
\noindent
where the constant $C$ is given by
\begin{eqnarray}
&&C=\frac{\hbar c}{32\pi^2}
\int_{0}^{\infty}\!\!\!y\,dy\int_{0}^{y}\!\!\!d\zeta
\left\{\ln\left[1-\left(
\frac{\alpha\pi y}{2\sqrt{f(\zeta,y)}+\alpha\pi y}
\right)^2\,e^{-y}\right]\right.
\nonumber \\
&&~~~~~~~~~~~
+\left.
\ln\left[1-\left(
\frac{\alpha\pi\sqrt{f(\zeta,y)}}{2y+\alpha\pi\sqrt{f(\zeta,y)}}
\right)^2\,e^{-y}\right]\right\}.
\label{eq11}
\end{eqnarray}
\noindent
Numerical computations of the integrals in Eq.~(\ref{eq11})
lead to
\begin{equation}
C=0.02101\frac{\hbar c}{32\pi^2}=
2.103\times 10^{-30}\,\mbox{J\,m}=
0.131\,\mbox{eV\,\AA}.
\label{eq12}
\end{equation}
\noindent
The result (\ref{eq12}) is in agreement with earlier
obtained\cite{16}
estimation $3C\sim 0.4\,$eV\,{\AA}
for the van der Waals pressure
in Eq.~(\ref{eq10}). In the next section the above results are
compared with other results obtained in the literature using the
Dirac model.

\section{Comparison of different results for dispersion interaction
between graphene sheets}

We begin with the van der Waals interaction between two graphene
sheets at zero temperature. In the first paper devoted to this
subject\cite{10} the van der Waals energy per unit area, as in
Eq.~(\ref{eq10}), was obtained with the coefficient $C$ equal to
$C=0.288\,$eV\,\AA. This is more than twice as large as our
value in Eq.~(\ref{eq12}).

In a later paper\cite{20} a smaller value for this coefficient
was obtained
$C=2.156\times 10^{-30}\,\mbox{J\,m}=0.134\,$eV\,{\AA}
in a
rather good agreement with our result (\ref{eq12}).
A slightly larger value computed\cite{20} might be explained
by slightly different value of the used Fermi velocity
($v_F=8.73723\times 10^{5}\,$m/s instead of
$v_F= 10^{6}\,$m/s as in our work). Although the formalism
used\cite{20} is nonretarded, a good agreement with our fully
relativistic computations was achieved. This again confirms
the conclusion\cite{16} that relativistic retardation does not
play a major role for graphene.

Now we discuss the computational results obtained at zero
temperature by using the models for conductivity of graphene
in terms of Lorentz-type oscillators.\cite{18}
By assuming that over a relatively wide range of photon
frequencies up to 3\,eV the graphene conductivity is
approximately constant equal to $\sigma_0=e^2/(4\hbar)$,
it was found\cite{18} that the Casimir pressure is given by the
second equality in Eq.~(\ref{eq10}). For the constant in this
equality it was derived\cite{18}
$3C=6.88\times 10^{-30}\,\mbox{J\,m}=0.430\,$eV\,{\AA}
leading to the following constant for the Casimir energy per
unit area
$C=2.29\times 10^{-30}\,\mbox{J\,m}=0.143\,$eV\,{\AA}.
This is a slightly larger value than was obtained in
Eq.~(\ref{eq12}) on the basis of the Dirac model.

\begin{figure}[b]
\vspace*{-13cm}
\centerline{\hspace*{1cm}
\includegraphics{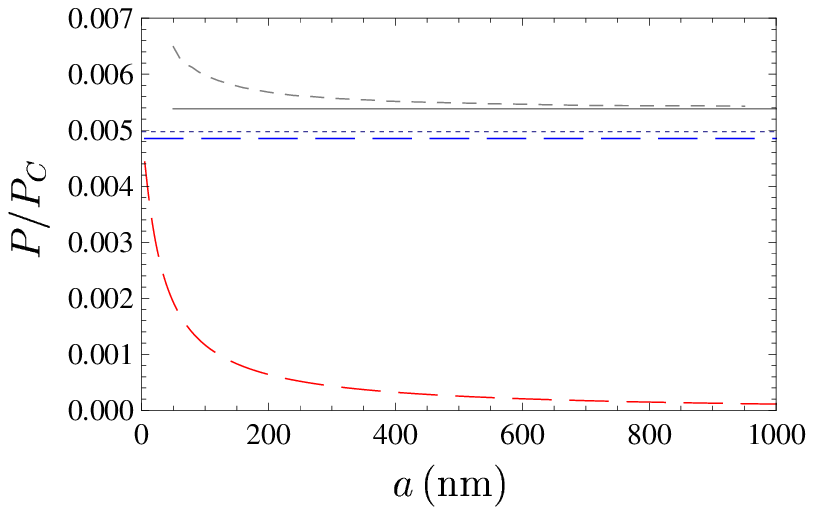}
}
\vspace*{-10cm}
\caption{\label{f4}(Color online)
The normalized van der Waals and Casimir pressure
of graphene-graphene interaction at $T=0\,$
as a function of separation.
The long-dashed lines show our results for
$\Delta=0.1\,$eV (bottom)  and 0\,eV (top).
The dotted line shows the results of Ref.~\cite{20}.
 The solid and dashed gray lines represent the results
of Ref.~\cite{18} for a constant and frequency-dependent
conductivity of graphene, respectively.
}
\end{figure}

The conductivity of graphene over a wider range of frequencies
was also modeled\cite{18} by assuming that its optical
properties are very similar to the in-plane optical properties
of graphite. The latter have been mapped to a series of Lorentz
oscillators with the Drude term over the frequency range from
0.1 to 40\,eV. As a result, the Casimir pressure between two
graphene sheets was computed using the Lifshitz formula.
The computational results normalized to the Casimir pressure
between two ideal metal planes taken from Fig.~4(b) in
Ref.~\cite{18} are plotted in our Fig.~\ref{f4} by the
gray short-dashed line. In the same figure, the gray solid
line shows the results obtained\cite{18} under an assumption
of constant graphene conductivity $\sigma_0$
($P/P_C=0.00529$).
As can be seen in Fig.~\ref{f4}, an assumption of the
frequency-dependent conductivity of graphene leads to significant
deviations at short separation distances. For the sake of
convenience in Fig.~\ref{f4} we also present the discussed above
results of Ref.~\cite{20} (the dotted line which corresponds to
$P/P_C=0.00497$) and our results for the gapless graphene sheets and
for graphene with the mass gap parameter $\Delta=0.1\,$eV
(they are shown by the top and bottom long-dashed lines,
respectively). For a gapless graphene it holds
$P/P_C=0.00485$.

It should be emphasized that although the formalism of reflection
coefficients expressed in terms of the polarization tensor
provides a reliable test for any alternative approach, at room
temperature the application region of the results computed at
zero temperature is restricted to only the shortest separations
below a few nanometers due to large thermal effects discussed in
Sec.~II. Because of this, below we compare our results with those
computed in the literature taking nonzero temperature into
account.

\begin{figure}[b]
\vspace*{-3cm}
\centerline{\hspace*{1cm}
\includegraphics{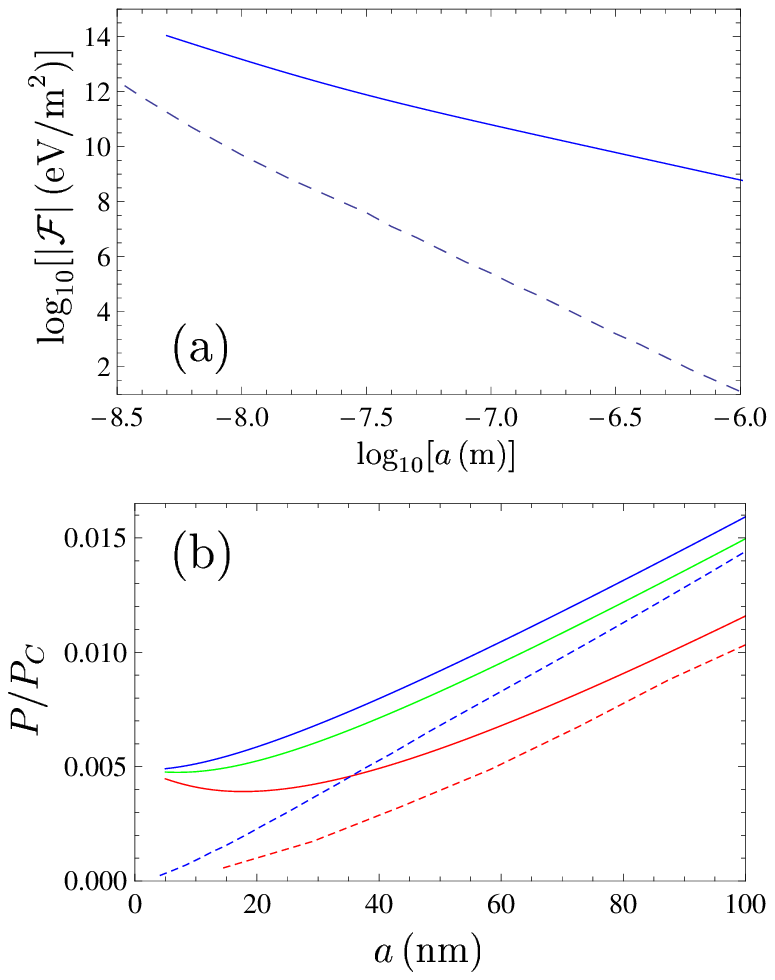}
}
\vspace*{-15cm}
\caption{\label{f5}(Color online)
Comparison of predictions at $T=300\,$K for 
(a) the van der Waals and Casimir free energy
per unit area where the solid and dashed lines represent
the results for pristine graphene obtained by us and
in Ref.~\cite{21}, respectively
(see Ref.~\cite{21} for possible explanations of
the discrepancy),
and
(b) the normalized van der Waals and Casimir pressure
where the solid lines represent our results
for $\Delta=0.1\,$eV, 0.05\,eV and 0\,eV from bottom
to top, respectively, and the  dashed lines show the
results of Ref.~\cite{22} for $\Delta=0.1\,$eV
(bottom) and $\Delta=0\,$eV (top).
}
\end{figure}

Using the Lifshitz theory and some version of the
response function of graphene in the random phase approximation,
the Casimir free energy between two graphene sheets was
computed\cite{21} within a wide range of separations at room
temperature $T=300\,$K. We show the computational results taken
from Fig.~4 of Ref.~\cite{21} in our Fig.~\ref{f5}(a) by the
dashed line for the pristine graphene.
For comparison purposes the solid line in
Fig.~\ref{f5}(a) reproduces our results for a pristine
graphene already shown by the top solid line in
Fig.~\ref{f1}(a) in another form.
As can be seen in Fig.~\ref{f5}(a), the dashed line deviates
significantly from the solid line obtained using the Dirac model
and the Lifshitz theory with the reflection coefficients
expressed via the polarization tensor (remind that the results
shown by the solid line are
in agreement with those obtained in Ref.~\cite{16}).
At asymptotically large separations the free energy of the
Casimir interaction is given by the zero-frequency term of the
Lifshitz formula (\ref{eq1}) and for two pristine graphene
sheets the formalism
using the temperature-independent response function of graphene
leads to ${\cal F}\sim 1/a^4$ (see Table~I in Ref.~\cite{21}),
whereas Eq.~(\ref{eq8}) demonstrates the
classical limit, ${\cal F}\sim -k_BT/a^2$,
as it should be at large separations (high temperatures).
According to Ref.~\cite{21} the reason for deviation from
the classical limit is that the used formalism\cite{21}
takes into account the direct temperature effects, as given
by the finite-temperature Lifshitz formula, but neglects the
indirect temperature effects arising from the temperature
dependence of the dielectric response of graphene (the latter
are taken into account by the polarization tensor at nonzero
temperature). Furthermore, Ref.~\cite{21} expects that for
graphene with high doping concentration the indirect
temperature effects at room temperature should be negligibly
small. Thus, for graphene with doping electron density
$10^{16}\,\mbox{m}^{-2}$, using the formalism accounting for
only the direct temperature effects it was found\cite{21}
that ${\cal F}\sim 1/a^2$ at separations $a>1\,\mu$m
in accordance to the classical limit.
{}From this it was concluded that for undoped graphene sheets
the formalism with neglected temperature dependence of the
dielectric response of graphene is applicable
only at vanishingly small separations of about a few
angstr\"{o}ms.\cite{21}
These suppositions of Ref.~\cite{21} concerning the origin
of disagreement between their results and the respective
results of Ref.~\cite{16}, coinciding with our results,
invite further investigation.

Now we consider computations of the thermal Casimir pressure
between two graphene sheets with both zero and nonzero mass
gap parameter.\cite{22} In this approach the dielectric
properties of graphene were described via the optical conductivity
calculated using the Kubo formalism. The computationsl
results\cite{22} for the van der Waals and Casimir pressure
normalized to the Casimir pressure between two ideal metal
planes are shown by the bottom and top dashed lines  in
Fig.~\ref{f5}(b) for graphene sheets with $\Delta=0.1\,$eV
and $\Delta=0\,$eV, respectively. These results are recalculated
from Fig.~2(a) in Ref.~\cite{22} by using their Eq.~(10).
For comparison purposes our results for the normalized
van der Waals and Casimir pressure are reproduced by the
three solid lines from bottom to top for $\Delta=0.1\,$eV,
0.05\,eV, and 0\,eV, respectively. As discussed in Sec.~II,
the top solid line is in agreement with respective
computational results of Ref.~\cite{16} where the case
of gapless graphene was also considered using an
alternative formalism. From  Fig.~\ref{f5}(b) it is
seen that the dashed lines deviate significantly from the
respective solid lines especially at short separations below
a few tens of nanometers. In the limiting case of large
separation distances (high temperatures) the used formalism
leads to one half of the result for two ideal metal
planes, i.e., the obtained Casimir pressure coincides with
the first term of our asymptotic expression (\ref{eq9}).
Thus, this formalism, although not enough precise at
moderate and short separations, satisfies the classical
limit.

\section{Conclusions and discussion}

In the foregoing we have investigated the van der Waals and
Casimir free energy and pressure between two graphene sheets
interacting via the zero-point and thermal fluctuations of
the electromagnetic field using the Dirac model of graphene.
This was done with the help of the Lifshitz theory where
the reflection coefficients were expressed via the components
of the polarization tensor in (2+1)-dimensional space-time.
In so doing both the pristine graphene and the gapped graphene
were considered.

It is common knowledge that graphene and graphene-based
nanostructures are the materials of high promise for many
prospective applications in micro- and nanoelectronics and,
more widely, in nanotechnology. Because of this, it is of high
priority to have reliable theoretical predictions for the
van der Waals and Casimir interactions between graphene sheets
and other carbon nanostructures spaced at separations below
a micrometer. At the present time, suspended graphene membranes
of sufficiently large area are already available.\cite{33}
It is highly probable that measurements of dispersion
interaction between graphene sheets and other carbon-based
nanostructures will be performed in the immediate future.
In this situation the reliable and confirmed theoretical results
for a simplest system, such as two graphene sheets, are
urgently needed.

The Lifshitz theory with reflection coefficients expressed
in terms of the polarization tensor provides a straightforward
formalism for the comparison with other approaches.
We have calculated the free energy and pressure
of dispersion interaction between two graphene sheets with zero
mass gap parameter and arrived to the results in agreement with
obtained earlier\cite{16} using Coulomb coupling between density
fluctuations. Specifically, the existence of large thermal effect
for two graphene sheets was confirmed as well as the asymptotic
behaviors of the free energy and pressure at short and large
separations. We have also generalized these results to the case
of graphene with a nonzero mass gap parameter. In this case the
thermal van der Waals and Casimir interactions between two
graphene sheets are shown to depend on an interrelation between
the temperature and the mass gap parameter.
The results obtained were compared with some other results in the
literature for graphene-graphene interaction at both zero and
nonzero temperature. This allowed to clarify the
regions of applicability
of several approaches to the definition of response function
and conductivity of graphene starting from the measure of
agreement between these approaches and the Lifshitz theory with
reflection coefficients found using the Dirac model.

In the future it would be topical to investigate the
van der Waals and Casimir interactions between
graphene sheets deposited on material substrates.
This subject is of much
interest for experiments in preparation. It is of interest also
to take into account deviations of the dispersion relation for
graphene quasiparticles from linearity at high energy.

\section*{Acknowledgments}

The authors are grateful to M.~Bordag for stimulating
discussions.



\begin{thebibliography}{99}
\bibitem {1}
J.~Mahanty and B.~W.~Ninham, {\it Dispersion Forces}
(Academic Press, London, 1976).
\bibitem{2}
M.~Kardar and R.~Golestanian,
Rev. Mod. Phys. {\bf 71}, 1233 (1999).
\bibitem{3}
V.~A.~Parsegian,
{\it Van der Waals Forces: A Handbook for Biologists,
Chemists, Engineers, and Physicists}
(Cambridge University Press, Cambridge, 2005).
\bibitem{4}
M.~Bordag, G.~L.~Klimchitskaya, U.\ Mohideen, and
V.\ M.\ Mostepanenko, {\it Advances in the Casimir Effect}
(Oxford University Press, Oxford, 2009).
\bibitem {5}
G.~L.~Klimchitskaya, U. Mohideen, and V.\ M.\ Mostepanenko,
 Rev. Mod. Phys. {\bf 81}, 1827 (2009).
 \bibitem{6}
A.~W.~Rodriguez, F.~Capasso, and S.~G.~Johnson,
Nature Photon. {\bf 5}, 211 (2011).
\bibitem{7}
G.~L.~Klimchitskaya, U. Mohideen, and V.\ M.\ Mostepanenko,
 Int. J. Mod. Phys. B {\bf 25}, 171 (2011).
\bibitem{8}
M.~S.~Dresselhaus,
Physica Status Solidi (b) {\bf 248}, 1566 (2011).
\bibitem{9}
A.~H.~Castro Neto, F.~Guinea, N.~M.~R.~Peres, K.~S.~Novoselov,
and A.~K.~Geim,
Rev. Mod. Phys. {\bf 81}, 109 (2009).
\bibitem{10}
J.~F.~Dobson, A.~White, and A.~Rubio,
{Phys. Rev. Lett.} {\bf 96}, 073201 (2006).
\bibitem{11}
I.~V.~Bondarev and Ph.~Lambin,
Phys. Rev. B {\bf 70}, 035407 (2004).
\bibitem{12}
E.~V.~Blagov, G.~L.~Klimchitskaya, and V.~M.~Mostepanenko,
{Phys. Rev. B} {\bf 71}, 235401 (2005).
\bibitem{13}
M.~Bordag,
J. Phys. A: Math. Gen. {\bf 39}, 6173 (2006).
\bibitem{14}
M.~Bordag, B.~Geyer, G.~L.~Klimchitskaya,
and V.~M.~Mostepanenko,
{Phys. Rev.} B {\bf 74}, 205431 (2006).
\bibitem{15}
M.~Bordag, I.~V.~Fialkovsky, D.~M.~Gitman, and
D.~V.~Vassilevich,
{Phys. Rev. B} {\bf 80}, 245406 (2009).
\bibitem{16}
G.~G\'{o}mez-Santos,
{Phys. Rev. B} {\bf 80}, 245424 (2009).
\bibitem{17}
I.~V.~Fialkovsky, V.~N.~Marachevsky, and
D.~V.~Vassilevich,
{Phys. Rev. B} {\bf 84}, 035446 (2011).
\bibitem{18}
D.~Drosdoff and L.~M.~Woods,
Phys. Rev. B {\bf 82}, 155459 (2010).
\bibitem {19}
D.~Drosdoff and L.~M.~Woods,
Phys. Rev. A {\bf 84}, 062501 (2011).
\bibitem{20}
B.~E.~Sernelius,
Europhys. Lett. {\bf 95}, 57003 (2011).
\bibitem{21}
J.~Sarabadani, A.~Naji, R.~Asgari, and R.~Podgornik,
Phys. Rev. B {\bf 84}, 155407 (2011).
\bibitem {22}
D.~Drosdoff, A.~D.~Phan, L.~M.~Woods, I.\ V.\ Bondarev,
and J.\ F.\ Dobson,
Eur. Phys. J. B {\bf 85}, 365 (2012).
\bibitem{23}
M.~Bordag, G.~L.~Klimchitskaya,
and V.~M.~Mostepanenko,
{Phys. Rev.} B {\bf 86}, 165429 (2012).
\bibitem{24}
Yu.~V.~Churkin, A.~B.~Fedortsov, G.~L.~Klimchitskaya,
and V.~A.~Yurova,
{Phys. Rev. B} {\bf 82}, 165433 (2010).
\bibitem{25}
T.~E.~Judd, R.~G.~Scott, A.~M.~Martin, B.~Kaczmarek,
and T.~M.~Fromhold,
New. J. Phys. {\bf 13}, 083020 (2011).
\bibitem{26}
M.~Chaichian, G.~L.~Klimchitskaya, V.\ M.\ Mostepanenko,
and A.~Tureanu,
Phys. Rev. A {\bf 86}, 012515 (2012).
\bibitem{27}
G.~Barton,
J. Phys. A {\bf 38}, 2997 (2005).
\bibitem{28}
S.~A.~Jafari,
J. Phys.: Cond. Mat. {\bf 24}, 205802 (2012).
\bibitem{29}
P.~K.~Pyatkovskiy,
J. Phys.: Cond. Mat. {\bf 21}, 025506 (2009).
\bibitem{30}
V.~P.~Gusynin, S.~G.~Sharapov, and J.~P.~Carbotte,
New J. Phys. {\bf 11}, 095013 (2009).
\bibitem{31}
V.~P.~Gusynin and S.~G.~Sharapov,
{Phys. Rev. B} {\bf 73}, 245411 (2006).
\bibitem{32}
J.~Feinberg, A.~Mann, and M.~Revzen,
Ann. Phys. (N.Y.) {\bf 288}, 103 (2001).
\bibitem{33}
B.~Alem\'{a}n, W.~Regan, S.~Aloni, V.~Altoe, N.~Alem,
C.~Girit, B.~Geng, L.~Maserati, M.~Crommie, F.~Wang,
and A.~Zettl,
ACS Nano {\bf 4}, 4762 (2010).
\end{thebibliography}
\end{document}